\documentclass[pra,twocolumn,showpacs,preprintnumbers,amsmath,amssymb]{revtex4}

\usepackage{amsmath,amssymb}
\usepackage{graphicx}
\usepackage{dcolumn}
\usepackage{bm}
\usepackage{times}

\begin{document}

\title{A simplified optical lattice clock}

\author{N. Poli, M. G. Tarallo\footnote{also at
Dipartimento di Fisica, Universit\`a di Pisa, Largo B. Pontecorvo
3 - 56100 Pisa}, M. Schioppo, C. W. Oates\footnote{permanent
address: National Institute of Standards and Technology, Boulder,
Colorado 80305 USA}, and G. M. Tino} \affiliation{Dipartimento di
Fisica and LENS, INFN-Sezione di Firenze, Universit\`a di Firenze,
Via Sansone,1 - 50019 Sesto Fiorentino - Italia}

\date{\today}

\begin{abstract}
Existing optical lattice clocks demonstrate a high level of
performance, but they remain complex experimental devices. In
order to address a wider range of applications including those
requiring transportable devices, it will be necessary to simplify
the laser systems and reduce the amount of support hardware. Here
we demonstrate two significant steps towards this goal:
demonstration of clock signals from a Sr lattice clock based
solely on semiconductor laser technology, and a method for finding
the clock transition (based on a coincidence in atomic
wavelengths) that removes the need for extensive frequency
metrology hardware. Moreover, the unexpected high contrast in the
signal revealed evidence of density dependent collisions in
$^{88}$Sr atoms.
\end{abstract}


\maketitle

\section{Introduction}

Optical atomic clocks have recently reached levels of performance
that are an order of magnitude or more beyond those of their
microwave counterparts, which have historically set the standards
for precision time/frequency metrology. With this improved level
of precision, scientists are performing new and more stringent
tests of fundamental physical principles such as relativity and
searches for temporal drifts in the fundament constants
\cite{rosenband08,blatt08}. Additionally, we anticipate that many
new applications will arise, including improved timing for space
travel and accelerator centers, as well as perhaps a new type of
relativity-based geodesy \cite{kleppner06}. These applications
will require more compact, robust, and versatile versions of
optical atomic clocks. Thus far, state-of-the-art optical clocks,
whether based on neutral or charged atoms, tend to be fairly
complicated pieces of scientific apparatus. They usually require
multiple (often sophisticated and power hungry) laser systems,
operate in well controlled environmental conditions, and rely on
links to well developed frequency standards. For these reasons,
most optical clocks are being developed in or near metrology
institutes.

Here we present a considerably simplified version of one of the
most promising optical clock systems, the Sr lattice clock. Due to
the favorable laser wavelengths required for its operation, the Sr
lattice clock is a good choice for streamlining, and in fact is
the most mature of lattice clocks, with many under development
around the world
\cite{ferrari03,Katori2,ludlow08,baillard08,legero07}. In a recent
demonstration, a Sr lattice clock achieved an absolute fractional
frequency uncertainty (for $^{87}$Sr) of 1.5 x 10$^{-16}$ and a
fractional instability of 10$^{-16}$ for an averaging time of
200~s \cite{ludlow08}. In our version of a $^{88}$Sr lattice
clock, we demonstrate two important simplifications. First, we use
only lasers that are based on semiconductor laser technology.
These lasers considerably simplify the apparatus and greatly
reduce the power/size requirements. Second, we take advantage of a
coincidence in the Sr atomic energy level system to greatly
simplify the needle-in-a-haystack challenge of locating the
ultranarrow spectroscopic feature used for stabilizing the clock
laser. This transfer cavity-based technique removes the need for
support from a standards institute or a GPS-referenced
femtosecond-laser frequency comb to calibrate the clock laser, and
can correct unexpected length changes in the clock reference
cavity (e.g., in a transportable device with a poorly controlled
environment). This simplified search method is further aided by a
new search algorithm based on chirping the laser frequency over a
one-second probe period, which has led to spectroscopic features
with a contrast greater than 99 $\%$. These developments represent
important steps towards a transportable device, which could enable
long-distance frequency comparisons at well below the 10$^{-15}$
level, and could also lead to simpler laboratory-based devices.

\section{Diode laser-based Sr lattice clock apparatus}

\begin{figure}
\begin{center}
\includegraphics[width=8.5 cm]{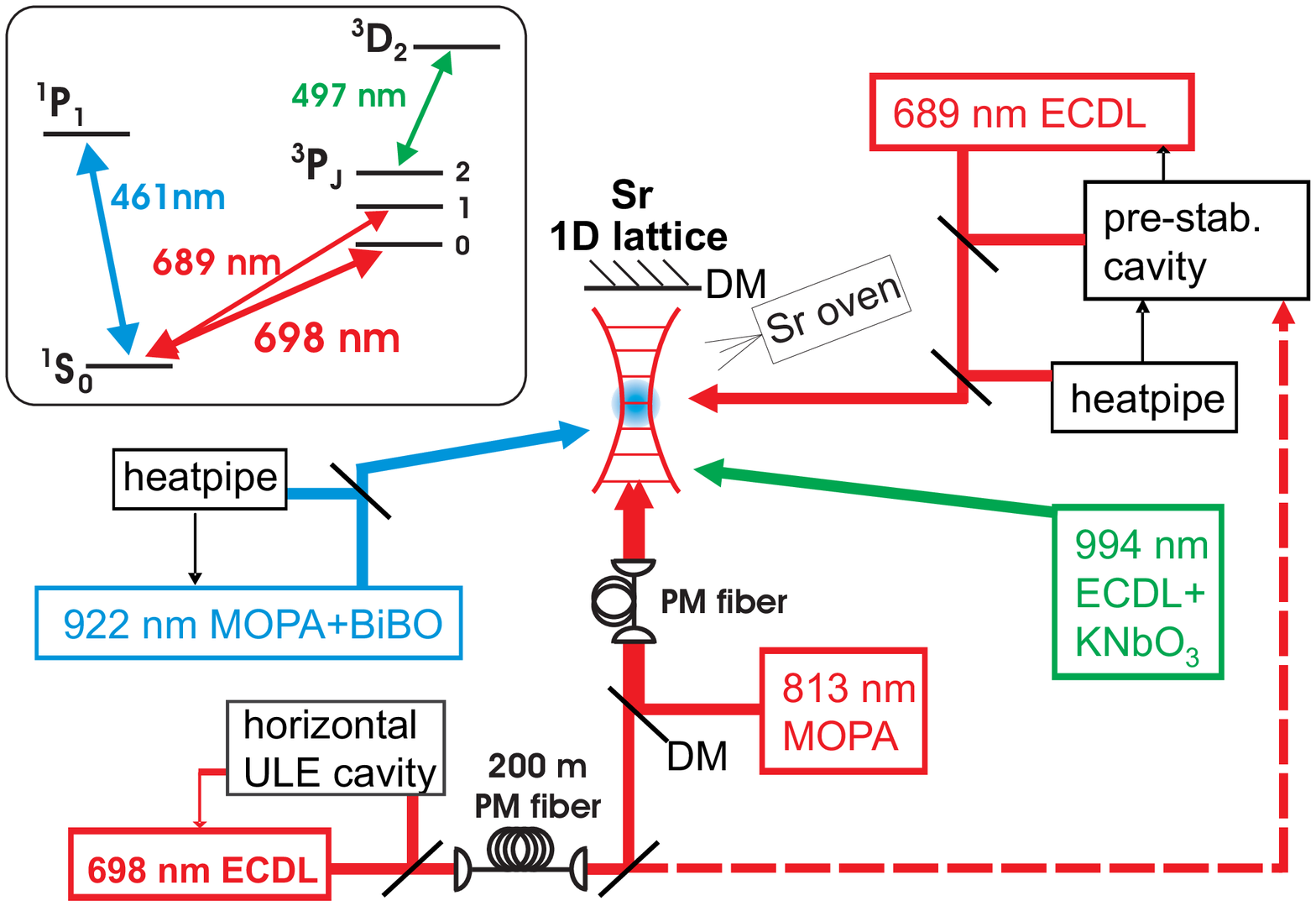}
\caption{Schematic of the Sr lattice clock apparatus. ECDL -
external cavity diode laser, MOPA - optical amplifiers that are
injection-locked by master lasers, DM - dichroic mirror. Inset:
relevant atomic levels and optical transitions for the Sr lattice
clock. All of the transitions are in the visible or in the near
infrared and can be reached with semiconductor laser
sources.\label{level}}
\end{center}
\end{figure}

As pointed out in the introduction, the level scheme of neutral Sr
offers some interesting advantages relative to other atoms/ions
when realizing a compact optical atomic clock. All of the relevant
transition wavelengths for cooling and trapping lie in the visible
or near infrared and can be easily reached with semiconductor
devices (see Fig.\ref{level}). Further reduction of the complexity
of the setup comes from the use of a magnetic-field-induced
spectroscopy scheme to observe the doubly-forbidden
$^1S_0\leftrightarrow\,^3P_0$ clock transition on bosonic isotopes
\cite{TY06,baillard07,poli08}. In this technique a static magnetic
field is used to mix the 3P0 metastable state with the 3P1 state,
enabling direct optical excitation of the clock transition with
tunable transition rates. Details on our experimental setup and
the laser sources used for cooling and trapping $^{88}$Sr have
been previously reported in \cite{poli07SPIE,poli07}. In brief,
$^{88}$Sr atoms are collected from a slowed Sr beam and cooled in
a two-stage magneto-optical trap (MOT). The MOT uses the strong
dipole allowed $^1S_0\leftrightarrow\,^1P_1$ transition at 461 nm
to cool and collect atoms, and then uses the intercombination
$^1S_0\leftrightarrow\,^3P_1$ transition at 689 nm to cool the
atoms to a temperature of 1~$\mu $K. To generate the 461 nm light
we double the frequency of a 922~nm diode/amplifier laser
combination with a resonant cavity containing a bismuth triborate
(BiBO) non-linear crystal. The second-stage cooling uses light
from a frequency-stabilized 689~nm diode laser that is locked to a
narrow resonance of a high-finesse Fabry-Perot cavity. The
frequencies of the trapping lasers are stabilized to their
respective atomic resonances by means of spectroscopy in Sr
heat-pipes. To increase the first-stage MOT lifetime (and thus,
the number of trapped atoms) repump light at 497~nm is used to
keep the atoms from getting shelved in the $^3P_2$ state by
exciting the $^3P_2\leftrightarrow\,^3D_2$ transition. The 497~nm
light is generated by doubling the frequency of a 994~nm diode
laser with a resonant cavity containing a KNbO$_3$ crystal.  To
further simplify the apparatus, this doubled laser system at
497~nm could be replaced with two semiconductor lasers at 707 nm
and 679~nm.

With this apparatus, typically $10^6$ atoms are trapped at
1~$\mu$K in a loading period of 300~ms. About $10^5$ atoms are
then loaded into a 1D standing wave optical lattice operating at
the magic wavelength for the $^1S_0\leftrightarrow\,^3P_0$ clock
transition (813.43~nm) \cite{Katori2}. A lattice potential depth
of about 8~$\mu$K (or 50 E$_r$ in units of the atomic recoil
energy, E$_r$) is realized by tightly focusing (waist $\sim$
30~$\mu$m) 200~mW of light coming from a 813~nm diode
laser/amplifier combination. To prevent optical feedback in the
tapered amplifier, which can induce phase fluctuations in the
lattice potential that lead to atom loss, two 40~dB stages of
optical isolation are necessary between the amplifier output and
the atoms. Under these conditions the measured lifetime in our
trap is around 2.5 s, limited by background gas collisions.

After the loading sequence, the $^1S_0\leftrightarrow\,^3P_0$
clock transition at 698~nm is probed after a 50~ms delay that
allows for the quadrupole magnetic MOT field to dissipate and the
static magnetic field to be turned on for the spectroscopy. To
reduce systematic effects related to the alignment of the probe
with respect to the infrared laser, the probe and lattice beams
are superimposed on a dichroic mirror and then coupled to the same
single-mode optical fiber that delivers the light to the atom
trap. After passing through the fiber, the two beams share the
same polarization optics that define a linear polarization aligned
with the vertical static magnetic field and the same achromatic
lens that focuses light on the atomic cloud. The red 698~nm clock
laser is produced in a separate quiet room by frequency
stabilization of a 698~nm diode laser to a high finesse ULE
(ultra-low expansion glass) cavity. The hertz-wide laser light is
then sent to the room containing the Sr apparatus through a 200~m-
long Doppler-cancelled optical fiber.

The frequency of the probe light from the master is shifted and
controlled with double-passed 200~MHz acusto-optic modulators
(AOM), and then amplified by an injection-locked diode laser. All
of the RF signals for driving the AOMs on the clock light are
generated with computer-controlled DDS (direct digital synthesis)
synthesizers phase-locked to a 10~MHz reference signal.  This
reference signal is generated by a BVA quartz crystal that is
slaved to a rubidium standard, which can be steered to a GPS clock
signal. The use of double-passed high-frequency AOMs and a
computer-controlled DDS system enables the possibility of chirping
the clock light frequency with a maximum span of 100~MHz (at
maximum rate of 2 MHz/s) without changing the optical path. We
could then scan over a large interval of frequency values in a
short period, which greatly accelerated initial searches for the
clock transition, as we will describe.  The atoms left in the
ground state after the spectroscopy pulse are detected and counted
by one of two methods: early measurements used calibrated
absorption of a blue resonant beam imaged on a CCD camera, while
high resolution spectrum are taken with a photomultiplier tube
that detects induced fluorescence (at 461 nm) from ground state
atoms.


\section{Finding the clock transition}

One challenge in optical clock research is to find an atomic
resonance with a natural (or induced) linewidth of below 10~mHz in
the continuum of frequencies around 10$^{15}$~Hz. Fortunately, the
problem is considerably simplified for resonances whose
frequencies have been previously precisely measured, as is the
case for the Sr and Yb lattice clocks \cite{poli08,boyd07}. But
even in this case, fairly precise calibration of the clock laser
is needed to find the transition, because the transition can
usually be artificially broadened only to tens of kilohertz, and
standard wavemeters have absolute uncertainties of 50~MHz to
100~MHz. Thus, with a simple point-per-second algorithm, the
search for the atomic transition could take hours. When the
transition is used on a fairly regular basis, sometimes after
extended periods of downtime, a more efficient search method is
needed.

Therefore, two separate problems have to be solved. First, the
transition must be initially found through a calibration of the
probe laser frequency. Second, a part of the apparatus has to then
be calibrated, so it can robustly be used to tune the probe laser
frequency quickly to the resonance on a day-to-day basis.

These problems are usually solved with a combination of a
femtosecond-laser frequency comb referenced to a frequency
standard (Cs fountain, GPS, etc.) and an ultrastable reference
cavity enclosed in a carefully monitored environment. The
standard-referenced comb can be used to provide a precise absolute
calibration of the probe laser frequency, while the cavity can
serve as a day-to-day or even a week-to-week reference. This
approach is feasible for systems located in metrology labs or labs
with access to metrology lab-based signals, however for a
transportable device, access to external standards could be
limited and perturbations to the cavity could be so large as to
move the reference fringe frequency by several megahertz, thereby
limiting its reliability as a day-to-day frequency reference.

A different approach to these challenges can be used in Sr by
taking advantage of the comparatively small difference in
wavelength between the second-stage cooling laser (689~nm) and the
clock laser (698~nm). To find the 698~nm transition for the first
time, it is possible to use the 689~nm light, which is locked to
the well-known frequency of the $^1S_0$$\leftrightarrow$$^3P_1$
transition \cite{ido05,tino94}, to calibrate a commercial 7-digit
wavemeter (Burleigh WP-1500). With this technique we were able to
measure the offset in the frequency reading of the wavemeter
($\sim$ 70~MHz), which we assumed would be similar for two
wavelengths so close together. In this way we could achieve an
absolute uncertainty for the 698~nm clock laser frequency of less
than 5~MHz.

\begin{figure}[t]\begin{center}
\includegraphics[width=7 cm]{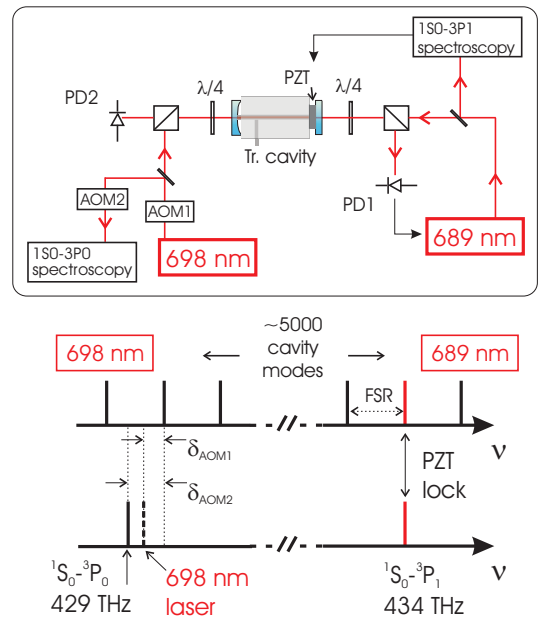}
\vspace{0.2 cm} \caption{Detail of the transfer cavity setup used
to obtain an optical frequency reference for the 698~nm clock
light in day-to-day operation. The two wavelengths (second stage
cooling at 689~nm and the clock transition at 698~nm) are coupled
with orthogonal polarization on the high finesse cavity locked to
the $^1S_0\leftrightarrow\,^3P_1$ transition. Acousto-optical
modulators (AOM1 and AOM2) are used to shift the clock laser
frequency so as to match the frequency of a resonant mode of the
transfer cavity and the atomic resonance, respectively. The
day-to-day drift of the clock laser frequency is then corrected
with AOM1 by monitoring the cavity transmission at 698~nm on PD1.
The free spectral range (FSR) of the transfer cavity is 1.5~GHz.
\label{trcavity}}
\end{center}
\end{figure}

While this method provides an uncertainty level low enough to find
the clock transition in a few trials (see next section), we can
remove the need for the wavemeter on a day-to-day basis
altogether, through use of the 689~nm pre-stabilization cavity as
a transfer cavity \cite{layer76,martin06}. Fig.~\ref{trcavity}
shows the experimental setup used to provide a repeatable
reference at 698~nm by means of the 689~nm pre-stabilization
cavity (made with a 10 cm long quartz spacer).

Its length is controlled by a piezo-electric transducer attached
to one of the cavity mirrors and is stabilized on long time scales
through spectroscopy on the $^1S_0\leftrightarrow\,^3P_1$
$^{88}$Sr transition. Thus a comb of resonant frequencies spaced
by the cavity free-spectral range is defined over tens of
nanometers. Light beams coming from the two sources at different
wavelengths are then coupled with orthogonal polarizations at
opposite ends of the cavity. Two acousto-optic modulators are used
to shift the frequency of the 698~nm light so that the first beam
is resonant with the $^1S_0\leftrightarrow\,^3P_0$ transition,
while the second is simultaneously resonant with a mode of the
transfer cavity stabilized on the $^1S_0\leftrightarrow\,^3P_1$
atomic transition.

We found that due to the small frequency difference between the
two wavelengths, they had comparable cavity finesse, with
resonance widths of 150~kHz. This value is sufficient to measure
and correct the possible drifts of the ULE cavity used for the
698~nm clock light with an accuracy of less than 10~kHz. We
verified the effectiveness of this approach when we were able to
recover the atomic signal quickly after a drift in the 698~nm
reference cavity of several megahertz occurred due to a large
variation in room temperature.

The discussion has thus far assumed that the transfer cavity
always locks to the same longitudinal cavity mode when it is
stabilized to the Sr heat pipe signal. In fact, the laser
sometimes locks to a nearby mode instead. In this case, the
frequency of the 698~nm light is no longer resonant with a mode of
the 689~nm-stabilized transfer cavity, due to a differential mode
shift. We have measured this shift and found it to be about 20~MHz
per mode. As long as the day-to-day shifts are less than 10~MHz
(they are usually less than 100~kHz in the lab), we can quickly
rectify the situation by either tuning the transfer cavity back to
the standard mode, or by accounting for the 20~MHz shift.

This method finds application not only in lab-based experiments,
but also in future transportable devices, where frequent
calibrations (e.g., due to shocks on reference cavities) could be
necessary, and where the access to frequency combs and frequency
references could be limited. By using the cavity transfer method,
even when shocks change the mode numbers considerably, a few
trials with different modes of the transfer cavity could be
sufficient to reset the offset between the transfer cavity and the
probe laser reference cavity without the need of external
references.

\section{Spectroscopy of the clock transition}

\begin{figure}[t]\begin{center}
\includegraphics[width=8 cm]{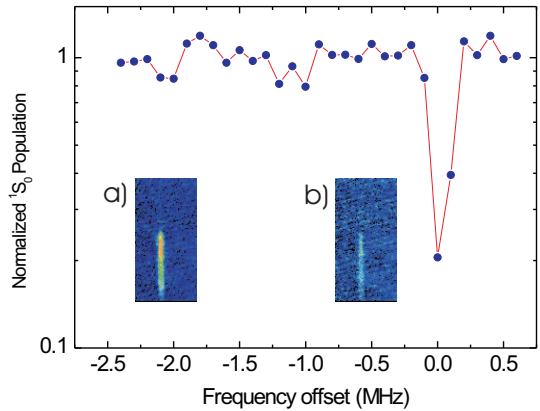}
\caption{Magnetic-field-induced spectroscopy spectrum of the clock
$^1S_0\leftrightarrow\,^3P_0$ transition of $^{88}$Sr trapped in
1D lattice at 698.4~nm, which has been observed with frequency
chirping of the clock light. Each point represents the number of
atoms left in the ground state after the 1~s excitation with
200~kHz chirping. Under these conditions, a period of about two
minutes is sufficient for acquiring the complete spectrum. In (a)
we show a typical CCD (charge-coupled device) camera absorption
image of the atoms left in the ground state with the probe laser
out of resonance (from which we determine the number of atoms),
while in (b) we show an image of the atoms after resonant
interaction with the clock laser. The spectrum has been taken with
the lattice wavelength detuned 0.4~nm from the magic wavelength, a
static magnetic field of 16~mT and a probe light intensity of
40~W/cm$^2$. The line contrast is greater than 80~\%, while no
repumping lasers are used to normalize the signal to number of
atoms. \label{clock1}}
\end{center}
\end{figure}

We excite the $^1S_0\leftrightarrow\,^3P_0$ clock transition in
$^{88}$Sr with the technique of magnetic field-induced
spectroscopy \cite{TY06}. We initially find the transition by
chirping the probe laser frequency over 200~kHz during each 1~s
probe pulse in order to increase the searching rate. With a
maximum magnetic field of 16~mT and a probe light intensity of
40~W/cm$^2$, the effective Rabi frequency is estimated to be
13~kHz. This frequency is sufficiently high to ensure observable
population transfer on resonance within the chirp range. Under
these conditions we can find the transition quickly by taking
200~kHz frequency steps around the resonance and measuring the
ground state population (see Figure \ref{clock1}). We prove this
search method to be effective even in the case of a poor knowledge
of the probe frequency, calibrated only using a commercial
wavemeter. Moreover, as we can see in Fig.~\ref{clock1}, the
search is greatly simplified because of the large contrast on the
signal. This enables us to find the transition even in the
presence of large fluctuations ($20~\%$) in atom number.

\begin{figure}[!t]\begin{center}
\includegraphics[width=8 cm]{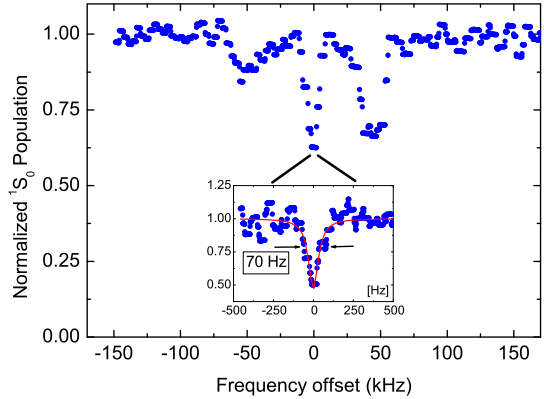}

\caption{High resolution spectrum of the
$^1S_0\leftrightarrow\,^3P_0$ transition taken by scanning the
probe laser frequency (without chirping). The spectrum is obtained
with a static magnetic field $B_0$~=~13~mT and probe intensity
$I$=38 W/cm$^2$ with an interaction period $\tau$~=~100~ms. The
separation between the carrier and the two sidebands
($\Delta$~=~50~kHz) indicates a lattice depth of 55 $E_r$. In the
inset is shown a scan of the carrier with $B_0$~=~1.3~mT,
$I$~=~6~W/cm$^2$ and the same interaction period. The observed 70
Hz linewidth is limited by the induced Rabi frequency in this
condition (see text). For both spectra the lattice wavelength has
been tuned near the magic wavelength at 813.428~nm\label{clock2}}
\end{center}
\end{figure}

In Fig.~\ref{clock2} the data were taken without frequency
chirping and with a reduced interaction time of 100~ms. The two
spectra have been taken with static magnetic fields ($B_0$) of
13~mT and 1.3~mT (for the spectrum in the inset) respectively, and
the probe intensities ($I$) were 38~W/cm$^2$ and 6~W/cm$^2$ (for
the inset), giving induced Rabi frequencies ($\Omega$) of 10~kHz
and 130~Hz (inset). To minimize the effects due to residual
frequency drifts in the clock laser, we modified the detection
scheme for the data shown in Fig.~\ref{clock2} to shorten the
measurement cycle duration to 1~s. We replaced the CCD with a
photomultiplier tube that collects fluorescence from ground-state
atoms that are excited by resonant light at 461~nm. Additionally,
we reduced the size of the residual drift of the clock frequency
that results from drifts in the ULE reference cavity. By feeding a
linear frequency ramp to the AOM located between the master clock
laser and the slave laser we have reduced this drift from standard
values of about 1~Hz/s to below 0.1~Hz/s.

Higher-resolution spectra reveal the usual set of three resonances
seen in lattice clock spectroscopy \cite{Katori2} (see
Fig.~\ref{clock2}). From the frequency spacing and the amplitude
ratio between the red and blue sidebands signals, we determined a
lattice depth of $U_0$~=~55~E$_r$ and an estimated temperature of
2.6~$\mu$K.
A clear difference in signal amplitude exists between spectra
taken with different interaction times, as seen in
Figs.~\ref{clock1} and \ref{clock2}. In particular, for long
interaction times with the probe light, we observed instances in
which the carrier and/or the blue sideband had virtually 100~$\%$
contrast (i.e., the ground -state population is completely
depleted). Under these overdriving conditions (pulse area $>$
1000$\pi$) we expect the amplitudes of the sidebands to be
considerably modified relative to the $\pi$ pulse case, but this
should not be true for the carrier. Normally, greatly overdriving
a transition leads to a maximum excitation level around 50~$\%$ as
atoms are driven back and forth between the excited and ground
states, as shown by the spectra taken with a shorter interaction
time in Fig.~\ref{clock2}. In fact, even for $\pi$-pulses used
with the carrier, 50~$\%$ percent is a typical excitation level
for lattice clock spectroscopy due to asymmetries in the
excitation for different atoms resulting from inhomogeneities for
quantities such as probe intensity or Rabi frequency. So the
depletion levels greater than 99~$\%$ were something of a
surprise, and these have shown a strong dependence on pulse
duration.

\begin{figure}[t]\begin{center}
\includegraphics[width=8 cm]{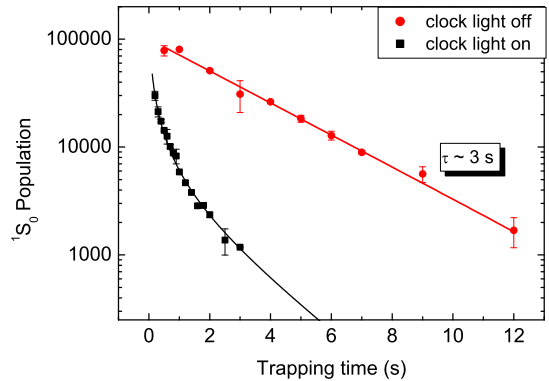}
\caption{Comparison of ground-state population decays. Data in
circles represent the lifetime of the 813 nm lattice trap and fit
well to a single exponential (solid line), vacuum limited to 2.9
s. The second, faster, decay is measured in the presence of the
clock light tuned to the carrier resonance at 698 nm. These data
agree well with a model based on a system of coupled rate
equations (see text) indicated by the solid (lower) line. The
non-exponential fast decay indicates an additional
density-dependent loss mechanism due to collisions of atoms in the
$^3$P$_0$ upper state with atoms in the ground state.
\label{decay}}
\end{center}
\end{figure}

To explore this effect, we examined the decay of the ground state
population for overdriven Rabi excitation (see Fig.~\ref{decay})
and compared it to the lifetime of the trap in the absence of
resonant excitation on clock transition. From the decay plots, it
is clear that when resonant clock light is present, the atoms are
lost at a higher rate compared to the normal decay rate that
results from background gas collisions that limit the trap
lifetime to about $\tau=1/\Gamma$=3~s. The faster decay in the
presence of excitation light can be well described by the solution
to a system of coupled rate equations $\dot{n}_S=-\Gamma\,
n_S-\Omega(n_S-n_P)-w\,n_S n_P$,
$\dot{n}_P=-\Gamma\,n_P-\beta\,n_P^2+\Omega(n_S-n_P)-w\,n_S n_P$
\cite{sterr08}. These equations describe the evolution of density
of atoms $n_S$ in the $^1$S$_0$ ground state and $n_P$ for the
excited $^3$P$_0$ state, respectively. In these two equations we
introduce additional density-dependent loss terms $\beta$ and $w$
associated with density-dependent losses due to collisions between
atom pairs in the excited state and collisions between ground- and
excited-state atoms, respectively. The atomic density in the
ground state is estimated from the measurement of the trap
frequencies and atom temperature with an uncertainty of 50~\%.
Assuming a value of $\beta$~=~4*10$^{-18}$~m$^3$/s
\cite{killian08} from the fit of the experimental data with the
solution of the decay equations, we estimate the value for
interstate collision coefficient to be $w$~=~3(2)*10$^{-17}$
m$^3$/s at measurement temperature of T~=~1.5~$\mu$K.

In one sense, this effect has the obvious benefit of making the
transition easy to find, even in the presence of reasonably large
shot-to-shot atom number fluctuations. Thus overdriving the
transition is now a standard part of our search algorithm.
However, this effect could also present a strong limitation for
operating a $^{88}$Sr 1D lattice clock at high densities, and
underfilled 2D traps or 3D traps could be the only way to overcome
this effect in view of making an accurate optical lattice clock
based on bosonic $^{88}$Sr.

\section{Summary}

As lattice clocks (and optical clocks in general) continue to
improve, it is also important that we continue to look for ways to
simplify the clocks, so they can reach a wider range of
applications. We have now demonstrated several key facets of an
ultimately transportable lattice clock system. We have constructed
an apparatus that uses only lasers based on semiconductor
technology, which greatly reduces the size and power requirements
over previous versions. We have also shown how the comparatively
small $^3$P fine structure splitting in Sr allows us to quickly
find the clock transition without relying on extensive metrology
apparatus. In the course of this development, we found a way to
achieve extremely high levels of ground-state population depletion
(more than 99 $\%$), which further expedites the search procedure.
An initial investigation into the source of the unexpectedly high
contrast revealed evidence of density dependent collisions, an
effect we plan to explore in more detail in future studies. We
thank G. Ferrari, A. Alberti, V. Ivanov, R. Drullinger and F.
Sorrentino for their work in the initial part of the experiment.
We thank S. Chepurov, D. Sutyrin, and U. Sterr for helpful
discussions and S. Xiao and R. Fox for a careful reading of the
manuscript. We thank INOA staff for the loaning of wavelength
meter used in this work. This work was supported by LENS under
contract RII3 CT 2003 506350, Cassa di Risparmio di Firenze,  ASI,
and ESA.


\begin{thebibliography}{22}

\bibitem{rosenband08}T. Rosenband et al., Science {\bf 319}, 1808 (2008)

\bibitem{blatt08} S. Blatt et al., Phys. Rev. Lett. {\bf 100}, 140801 (2008)

\bibitem{kleppner06} D. Kleppner, Phys. Today {\bf 59} , 10 (2006)

\bibitem{ferrari03} G. Ferrari et al., Phys. Rev. Lett. {\bf 91}, 243002 (2003)

\bibitem{Katori2} M. Takamoto, F.-L. Hong, R. Higashi, and H. Katori, Nature (London) {\bf 435}, 321 (2005).

\bibitem{ludlow08} A. D. Ludlow et al., Science {\bf 319}, 1805 (2008)

\bibitem{baillard08} X. Baillard et al., Eur. Phys. Jour. D {\bf
48}, 11 (2008).

\bibitem{legero07} T. Legero, J.S.R. Winfred, F. Riehle, U. Sterr,
Frequency Control Symposium, 2007 Joint with the 21st European
Frequency and Time Forum. IEEE International, 119 (2007)

\bibitem{TY06} A.V. Taichenachev, V.I.
Yudin, C.W. Oates, C.W. Hoyt, Z.W. Barber, and L. Hollberg, Phys.
Rev. Lett. {\bf 96}, 083001 (2006).

\bibitem{baillard07} X. Baillard et al., Opt.
Lett. {\bf 32}, 1812 (2007).

\bibitem{poli08}  N. Poli  et al.,  Phys. Rev. A {\bf 77}, 050501 (2008)

\bibitem{poli07SPIE} N. Poli et al., Proc. SPIE {\bf 6673}, 66730F
(2007)

\bibitem{poli07} N. Poli, R. E.  Drullinger, M. G. Tarallo, and G. M. Tino,
Frequency Control Symposium, 2007 Joint with the 21st European
Frequency and Time Forum. IEEE International, 655, (2007)

\bibitem{boyd07} M. M. Boyd, A. D. Ludlow, S. Blatt, S. M. Foreman, T. Ido, T. Zelevinsky, and J.
Ye, Phys. Rev. Lett. {\bf 98}, 083002 (2007).

\bibitem{ido05} T. Ido, T. H. Loftus, M. M. Boyd,
A. D. Ludlow, K. W. Holman, and J. Ye Phys. Rev Lett {\bf 94},
153001 (2005)

\bibitem{tino94} G.M. Tino \emph{et al.}, Appl. Phys. B {\bf 55}, 397 (1994)

\bibitem{layer76} H. P. Layer, R. D. Deslattes, and W. G. Schweitzer, Jr., Appl. Opt. {\bf 15}, 734 (1976)

\bibitem{martin06} P. Bohlouli-Zanjani, K. Afrousheh, and J. D. D. Martin, Rev. Sci. Instrum. {\bf 77}, 093105 (2006)

\bibitem{sterr08} Similar set of equations have also been used in J. S. R. Vellore Winfred et al., Proc. of 7$^{th}$ Symposium of Frequency standards and Metrology (2008)

\bibitem{killian08} A. Traverso et al. arXiv:0809.0936 (2008)









\end{thebibliography}
\end{document}